\documentclass[nofootinbib,prd,twocolumn,showpacs,showkeys,preprintnumbers]{revtex4-1}
\usepackage{hyperref,amssymb,amsmath,mathrsfs,bm,graphicx}
\begin{document}

\title {Dynamics of hyperbolically symmetric fluids}
\author{L. Herrera}
\email{lherrera@usal.es}
\affiliation{Instituto Universitario de F\'isica
Fundamental y Matem\'aticas, Universidad de Salamanca, Salamanca 37007, Spain}
\author{A. Di Prisco}
\email{alicia.diprisco@ciens.ucv.ve; adiprisc@fisica.ciens.ucv.ve}
\affiliation{Escuela de F\'\i sica, Facultad de Ciencias, Universidad Central de Venezuela, Caracas 1050, Venezuela}
\author{J. Ospino}
\email{j.ospino@usal.es}
\affiliation{Departamento de Matem\'atica Aplicada and Instituto Universitario de F\'isica
Fundamental y Matem\'aticas, Universidad de Salamanca, Salamanca 37007, Spain}
\begin{abstract}
We  study the general properties of dissipative  fluid distributions endowed with hyperbolical symmetry. Their
physical properties are analyzed in detail. It is shown that  the energy density is necessarily negative and the fluid distribution cannot fill the region close to the
center of symmetry. Such a region may be represented by a vacuum cavity around the center. By assuming a causal  transport equation some interesting thermodynamical properties of these fluids are found. Several exact analytical solutions  which evolve in the quasi--homologous regime and satisfy the vanishing complexity factor condition, are exhibited
\end{abstract}
\maketitle

\section{Introduction}
In  a recent  paper  \cite{st1}, a general study on the properties of static fluid distributions endowed with hyperbolical symmetry was carried out.  The main motivation  (but not the only one) behind  such endeavor  was the necessity to provide a rigorous description of fluid distributions sourcing the line element
\begin{eqnarray}
ds^2&=&-\left(\frac{2M}{R}-1\right)dt^2+\frac{dR^2}{\left(\frac{2M}{R}-1\right)}+R^2d\Omega^2, \nonumber \\
d\Omega^2&=&d\theta^2+\sinh^2 \theta d\phi^2,
\label{w3}
\end{eqnarray}
which in its turn is assumed to be the line element at the interior of the horizon, proposed  in \cite{1, 2} as an alternative  global description of the  Schwarzschild black hole.

Such a proposal  is motivated by    the  fact that it is impossible to remove the coordinate singularity in the line element, keeping at the same time the static form of the
Schwarzschild metric (in the whole space--time) 
\cite{rosen}. Thus, the regular extension of the Schwarzschild metric to the whole space--time  may be achieved but at the price to admit a non-static space--time inside the horizon \cite{Rin, Caroll}.

Then, from the belief  that  any dynamic regime should eventually lead to an equilibrium final state,
a static solution  has to be expected in the whole space--time.

Accordingly,  the   model proposed in \cite{1} describes the space time as
consisting of two four-dimensional manifolds, the outer one   described by  the usual  Schwarzschild metric on the exterior side of the horizon
and the inner one described by (\ref{w3}).  A change in signature as well
as a  change in the symmetry at the horizon are required.

The metric (\ref{w3}) is a static solution admitting the four Killing vectors

\begin{equation}
\mathbf{K}_{(\mathbf{0})} = \partial _{\mathbf{t}},
\label{sest}
\end{equation}
and
\begin{eqnarray}
 {\bf K_{(2)}}=-\cos \phi
\partial_{\theta}+\coth\theta \sin\phi \partial_{\phi},\nonumber \\
{\bf K_{(1)}}=\partial_{\phi}, \quad {\bf K_{(3)}}=\sin \phi \partial_{\theta}+\coth\theta \cos\phi
\partial_{\phi}.
\label{shy}
\end{eqnarray}

Solutions to the Einstein equations  endowed with the hyperbolic symmetry
(\ref{shy}) has been  the subject of research  in
different contexts (see \cite{Ha, 1n, Ga, Ri, mim, Ka, Ma,  mimII} and references therein).

Since the fluid that sources the line element (\ref{w3}) is considered as the final state ensuing from a dynamical regime, the obvious question is: What are the general properties of the fluid distribution during this evolving regime, before reaching the equilibrium?

Our purpose in this work is to answer to the above question by
 carrying  on  a comprehensive study on the physical properties of  evolving fluid distributions in the region inner to the
horizon, endowed with the hyperbolical symmetry (\ref{shy})  and that eventually may converge to the static fluid distributions described in \cite{st1}.

 We shall deploy all required equations for a full description of the fluid distribution, including a transport equation. Some specific analytical solutions to these equations will be exhibited. The solutions will be obtained assuming the quasi--homologous condition for their evolution, and the vanishing of the complexity factor.

 As we shall see below within the region $r<2m$, where $m(t,r)$ is a suitable definition of the mass function,  the
energy density is negative, and the central region cannot be filled with  our fluid distribution. Thus either the center is surrounded by an empty cavity or by a fluid distribution not endowed with hyperbolical symmetry. A discussion about the physical meaning of the obtained results  is presented.
\section{The general setup of the problem: notation, variables and equations}
We consider  hyperbolically symmetric  distributions  of evolving
fluids, which may be  bounded  from outside by a surface  $\Sigma^{e}$, and, in the case when   a cavity  is present, are also bounded from inside by a  surface   $\Sigma^{i}$.  The fluid is assumed to be locally anisotropic (principal stresses unequal) and undergoing dissipation in the form of heat flow (diffusion approximation).

Choosing comoving coordinates, the general
interior metric can be written as
\begin{equation}
ds^2=-A^2dt^2+B^2dr^2+R^2(d\theta^2+\sinh^2\theta d\phi^2),
\label{1}
\end{equation}
where  $A$, $B$ and $R$ are assumed
positive, and  due to the symmetry (\ref{shy}) are functions of $t$ and $r$. We number the coordinates $x^0=t$, $x^1=r$, $x^2=\theta$
and $x^3=\phi$.  $A$ and $B$ are dimensionless, whereas $R$ has the same dimension as $r$.

The energy momentum tensor $T_{\alpha\beta}$ of the fluid distribution
may be written as
\begin{eqnarray}
T_{\alpha\beta}&=&(\mu +
P_{\perp})V_{\alpha}V_{\beta}+P_{\perp}g_{\alpha\beta}+(P_r-P_{\perp})\chi_{
\alpha}\chi_{\beta}\nonumber \\&+&q_{\alpha}V_{\beta}+V_{\alpha}q_{\beta}
, \label{3}
\end{eqnarray}
where $\mu$,  $P_r$, 
$P_{\perp}$,  $q^{\alpha}$, $V^{\alpha}$ have the usual meaning, 
and $\chi^{\alpha}$ is unit four--vector along the radial direction.  Besides, the four--vectors $V^{\alpha}$, $q^\alpha$   and $\chi^{\alpha}$ 
satisfy
\begin{eqnarray}
V^{\alpha}V_{\alpha}=-1, \;\; V^{\alpha}q_{\alpha}=0, \;\; \chi^{\alpha}\chi_{\alpha}=1,\;\;
\chi^{\alpha}V_{\alpha}=0.
\end{eqnarray}

 Since the Lie derivative and the partial derivative commute, then
\begin{equation}
\mathcal{L}_\chi (R_{\alpha \beta}-\frac{1}{2}g_{\alpha \beta}{\cal R})=8\pi \mathcal{L}_\chi T_{\alpha \beta}=0,
\label{ccm1}
\end{equation}
implying because of (\ref{shy}) that all physical variables only depend on $t$ and $r$.

It will be convenient to express the  energy momentum tensor  (\ref{3})  in the equivalent (canonical) form
\begin{equation}
T_{\alpha \beta} = {\mu} V_\alpha V_\beta + P h_{\alpha \beta} + \Pi_{\alpha \beta} +
q \left(V_\alpha \chi_\beta + \chi_\alpha V_\beta\right) \label{Tab}
\end{equation}
with
$$ P=\frac{P_{r}+2P_{\bot}}{3}, \qquad h_{\alpha \beta}=g_{\alpha \beta}+V_\alpha V_\beta,$$

$$\Pi_{\alpha \beta}=\Pi\left(\chi_\alpha \chi_\beta - \frac{1}{3} h_{\alpha \beta}\right), \qquad
\Pi=P_{r}-P_{\bot}.$$

Since we are considering comoving observers, we have
\begin{eqnarray}
V^{\alpha}&=&A^{-1}\delta_0^{\alpha}, \;\;
q^{\alpha}=qB^{-1}\delta^{\alpha}_1, \;\;
\chi^{\alpha}=B^{-1}\delta^{\alpha}_1.
\end{eqnarray}

It is worth noticing that  bulk or shear viscosity could be introduced by redefining 
 the radial and tangential pressures.  In addition, dissipation in the free
streaming approximation can be absorbed in $\mu, P_r$ and $q$.

\subsection{Einstein equations and conservation laws}

The Einstein equations for (\ref{1}) and (\ref{Tab}), are
\begin{eqnarray}
  8\pi \mu&=& -\frac{1}{R^2}-\frac{1}{B^2}\left
  [-\frac{2B^\prime}{B}\frac{R^\prime}{R}+\left(\frac{R^\prime}{R}\right)^2+\frac{2R^{\prime\prime}}{R} \right ]\nonumber\\
   &+& \frac{1}{A^2}\left ( \frac{2\dot B}{B}\frac{\dot R}{R}+\frac{\dot R^2}{R^2} \right ),\label{EE00}
\end{eqnarray}

\begin{equation}\label{EE01}
 4\pi q=-\frac{1}{AB}\left ( \frac{R^\prime }{R}\frac{\dot{B}}{B}+\frac{A^\prime}{A}\frac{\dot
 R}{R}-\frac{\dot{R}^\prime}{R}\right),
\end{equation}

\begin{eqnarray}
8\pi P_r &=& \frac{1}{R^2}+\frac{1}{B^2}\left [\frac{2A^\prime}{A}\frac{R^\prime}{R}+\left(\frac{R^\prime}{R}\right)^2 \right
]\nonumber \\
   &+& \frac{1}{A^2}\left ( \frac{2\dot{A}}{A}\frac{\dot{R}}{R}-\frac{\dot R^2}{R^2}-\frac{2\ddot{R}}{R}\right
   ),\label{EE11}
\end{eqnarray}

\begin{eqnarray}
 8\pi P_\bot&=& \frac{1}{B^2}\left (
 -\frac{A^\prime}{A}\frac{B^\prime}{B}+\frac{A^\prime}{A}\frac{R^\prime}{R}-\frac{B^\prime}{B}\frac{R^\prime}{R}
 +\frac{A^{\prime\prime}}{A}+\frac{R^{\prime\prime}}{R} \right ) \nonumber\\
   &+& \frac{1}{A^2}\left (
   \frac{\dot{A}}{A}\frac{\dot{B}}{B}+\frac{\dot{A}}{A}\frac{\dot{R}}{R}-\frac{\dot{B}}{B}\frac{\dot{R}}{R}
   -\frac{\ddot B}{B}-\frac{\ddot R}{R}  \right ),\label{EE22}
\end{eqnarray}
where dots and primes denote derivative with respect to $t$ and $r$ respectively.
It is worth noticing the difference between these equations and the corresponding to the spherically symmetric case (see for example Eqs.(7)--(10) in \cite{epjc}).

The conservation laws $T^{\mu \nu}_{;\mu}=0$, as in the spherically symmetric case,  have only two independent components, which are displayed in Appendix A.
\subsection{Kinematical variables}

The four--acceleration $a_{\alpha}$ and the expansion $\Theta$ of the fluid are
given by
\begin{equation}
a_{\alpha}=V_{\alpha ;\beta}V^{\beta}, \;\;
\Theta={V^{\alpha}}_{;\alpha}. \label{4b}
\end{equation}
From  which we obtain  for the  four--acceleration  and its scalar $a$,
\begin{equation}
a_1=\frac{A^{\prime}}{A}, \;\; a=\sqrt{a^{\alpha}a_{\alpha}}=\frac{A^{\prime}}{AB}\Rightarrow a^\alpha=a\chi^\alpha, \label{5c}
\end{equation}
and for the expansion
\begin{equation}
\Theta=\frac{1}{A}\left(\frac{\dot{B}}{B}+2\frac{\dot{R}}{R}\right).
\label{5c1}
\end{equation}

The   shear tensor $\sigma_{\alpha\beta}$ is defined by (the vorticity vanishes identically)
\begin{equation}
\sigma_{\alpha\beta}=V_{(\alpha
;\beta)}+a_{(\alpha}V_{\beta)}-\frac{1}{3}\Theta h_{\alpha\beta},
\label{4a}
\end{equation}
its non zero components are
\begin{equation}
\sigma_{11}=\frac{2}{3}B^2\sigma, \;\;
\sigma_{22}=\frac{\sigma_{33}}{\sinh^2\theta}=-\frac{1}{3}R^2\sigma,
 \label{5a}
\end{equation}
and its scalar
\begin{equation}
\frac{3}{2}\sigma^{\alpha\beta}\sigma_{\alpha\beta}=\sigma^2,
\label{5b}
\end{equation}
reads
\begin{equation}
\sigma=\frac{1}{A}\left(\frac{\dot{B}}{B}-\frac{\dot{R}}{R}\right).\label{5b1}
\end{equation}

All the expressions above are the same, in terms of the metric functions, as in the spherically symmetric case.
\subsection{The Weyl tensor}
Using Maple we may easily obtain the Weyl tensor corresponding to our metric (\ref{1}). Thus, the magnetic part of the Weyl tensor vanishes, whereas  its electric part may be written as

\begin{equation}\label{weyl1}
  E_{\alpha\beta}=\mathcal{E}\left(\chi_\alpha \chi_\beta-\frac{1}{3}h_{\alpha\beta}\right),
\end{equation}
with
\begin{widetext}
\begin{eqnarray}
\mathcal{E}&=&\frac{1}{2B^2}\left[
-\frac{A^\prime}{A}\frac{R^\prime}{R}-\frac{A^\prime}{A}\frac{B^\prime}{B}+\frac{R^\prime}{R}\frac{B^\prime}{B}+
\left(\frac{R^\prime}{R}\right)^2+\frac{A^{\prime \prime}}{A}-\frac{R^{\prime \prime}}{R}\right]\nonumber \\
&+& \frac{1}{2A^2}\left[ \frac{\dot A}{A}\frac{\dot B}{B}-\frac{\dot A}{A}\frac{\dot R}{R}+\frac{\dot
R}{R}\frac{\dot B}{B}-
\left(\frac{\dot R}{R}\right)^2-\frac{\ddot B}{B}+\frac{\ddot R}{R}\right]+\frac{1}{2R^2}.\label{escE}
\end{eqnarray}
\end{widetext}

\subsection{The mass function}
Following \cite{Misner}  we may define the mass function  as
\begin{equation}
m(r,t)=-\frac{R}{2}R^{3}_{232}=\frac{R}{2}\left[\left(\frac{R^{\prime}}{B}\right)^2-\left(\frac{\dot
R}{A}\right)^2+1\right],
 \label{fmasa}
\end{equation}
where the Riemann tensor component $R^{3}_{232}$ is now calculated for (\ref{1}).

Introducing the proper time derivative $D_T$, and the proper radial derivative $D_R$  by
\begin{equation}
D_T=\frac{1}{A}\frac{\partial}{\partial t}, \label{16}
\end{equation}

 \begin{equation}
D_R=\frac{1}{R^{\prime}}\frac{\partial}{\partial r},\label{23a}
\end{equation}
we can define the velocity $U$ as
\begin{equation}
U=D_TR, \label{19}
\end{equation}
which must be smaller than $1$ (in relativistic units).

Indeed, in Gaussian coordinates, the position of each fluid element may be given as
\begin{equation}
x^\alpha=x^\alpha(y^a,s),
\label{vel1}
\end{equation}
where $s$ is the proper time along the world line of the particle, and $y^a$ (with $a$ running from 1 to 3) is the position of the particle on a three-dimensional hypersurface (say $\Sigma$).

Next, for an infinitesimal variation of the world line we have
\begin{equation}
\delta x^\alpha=\frac{\partial x^\alpha}{\partial y^a} \delta y^a,
\label{vel3}
\end{equation}
from which it follows
\begin{equation}
D_T (\delta x^\alpha)=V^\alpha_{;\beta}\delta x^\beta.
\label{vel4}
\end{equation}
We can define the position vector of the particle $y^a+\delta y^a$ relative to the particle $y^a$ on $\Sigma$, as
\begin{equation}
\delta_{\bot}x^\alpha=h^\alpha_{\beta} \delta x^\beta.
\label{vel6}
\end{equation}
Then the relative velocity between these two particles, is
\begin{equation}
u^\alpha=h^\alpha_{\beta} D_T(\delta_{\bot} x^\beta),
\label{vel7}
\end{equation}
and considering (\ref{vel4}) and  (\ref{vel6}) it follows that
\begin{equation}
u^\alpha=V^\alpha_{;\beta} \delta_{\bot} x^\beta.
\label{vel8}
\end{equation}

Defining the infinitesimal distance between two neighboring points on $\Sigma$ by
\begin{equation}
\delta l^2=g_{\alpha \beta} \delta_{\bot} x^\beta \delta_{\bot} x^\alpha,
\label{vel9}
\end{equation}
then it can be shown (see \cite{wang} for details) that 
\begin{equation}
\delta l D_T(\delta l)= \delta_{\bot} x^\beta \delta_{\bot} x^\alpha\left(\sigma_{\alpha \beta}+\frac{1}{3}h_{\alpha \beta} \Theta\right),
\label{vel13}
\end{equation}
or, introducing the spacelike unit vector
\begin{equation}
e^\alpha=\frac{\delta_{\bot} x^\alpha}{\delta l}, \label{unit}
\end{equation}

\begin{equation}
 \frac{D_T(\delta l)}{\delta l}= e^\alpha e^\beta \sigma_{\alpha \beta}+\frac{\Theta}{3}.
\label{vel14}
\end{equation}

The above expressions are completely general, let us now consider our hyperbolically symmetric line element  and apply (\ref{vel14}) to two neighbouring points on a closed curve ($S$) along the $\phi$ direction ($r=constant; \theta=constant$). In this case we have $e^\alpha \equiv(0, 0, 0, \frac{1}{R \sinh\theta})$
and using (\ref{5c1}), (\ref{5a}) and   (\ref{5b1}) in  (\ref{vel14}) we obtain
\begin{equation}
 \frac{D_T(\delta l)}{\delta l}= \frac{\dot R}{AR}=\frac{U}{R}.
\label{vel16}
\end{equation}
Now, the $D_T(\delta l)$ above is the relative velocity between two neighbouring points on $S$. This quantity of course must be smaller than one (in relativistic units). On the other hand the rate of variation of the total length ($L$)  of $S$  per unit of proper time (say $V_L$)  is also a velocity, and thereof must be smaller than $1$ and because of the axial symmetry it  is just the sum of  (\ref{vel16}) over all the curve $S$. Thus we have  
\begin{equation}
\frac{R}{L}V_L= U.
\label{vel17}
\end{equation}

For any value of $\theta$,  $\frac{R}{L}<1$ and thereof $U<1$. 

Then, since $U <1$, it follows at once from (\ref{fmasa}) that $m$ is a positive defined quantity.
Also, (\ref{fmasa}) can be rewritten as
\begin{equation}
E \equiv \frac{R^{\prime}}{B}=\left(\frac{2m}{R}+U^2-1\right)^{1/2}.
\label{20x}
\end{equation}

Using (\ref{fmasa})  with (\ref{16}) and (\ref{23a}) we obtain

\begin{eqnarray}
D_Tm=4\pi\left(P_rU+ qE\right)R^2,
\label{22Dt}
\end{eqnarray}
and
\begin{eqnarray}
D_Rm=-4\pi\left(\mu+q\frac{U}{E}\right)R^2,
\label{27Dr}
\end{eqnarray}
which implies
\begin{equation}
m=-4\pi\int^{r}_{0}\left( \mu +q\frac{U}{E}\right)R^2R^\prime dr, \label{27intcopy}
\end{equation}
satisfying the regular condition  $m(t,0)=0$.

Integrating (\ref{27intcopy}) we find
\begin{equation}
\frac{3m}{R^3} = -4\pi {\mu} +\frac{4\pi}{R^3} \int^r_0{R^3\left(D_R{ \mu}-3 q \frac{U}{RE}\right) R^\prime dr}.
\label{3m/R3}
\end{equation}

Since any causal transport equation is based on the assumption that the fluid is not very far from thermal equilibrium then $q<<\vert\mu\vert$. This implies from (\ref{27intcopy}) that $\mu$ is necessarily negative, if we assume   the condition $R^\prime>0$ to avoid shell crossing,  and remind that $m>0$ and $E$ is a regular function within the fluid distribution.

 Furthermore, it follows from (\ref{27intcopy}) that whenever the energy density  is regular, then $m\sim r^3$ as $r$ tends to zero. However, in this same limit $U\sim 0$, and $R\sim r$ implying because of (\ref{20x}) that the central region  cannot be filled with our fluid distribution. Among the many  possible scenarios we shall assume here that the center is surrounded by a vacuum cavity. However, it should be clear that this is just one of the possible choices, which even if having implications on  specific models, does not affect  the general properties of the  fluids endowed with hyperbolical symmetry.

 The two above mentioned features of the fluid appear also in the static case \cite{st1}.

Before concluding this section it is worth  discussing with some detail  on equation (\ref{pprimaU2}), and compare it with the corresponding equation for the spherically symmetric case (see eq.(C6) in \cite{epjc}).

 First of all let us notice that it has  the ``Newtonian'' form
$Force = Mass \ density \times \ Acceleration$.  Let us next analyze the different terms in (\ref{pprimaU2}). The first term on the right represents the gravitational interaction,  it is the product of the passive gravitational mass density (p.g.m.d) ($\mu+P_r$), which due to the fact that the energy density is negative, would be negative,  and the active gravitational mass (a.g.m) ($4\pi P_r R^3-m$)  which would also be negative for most equations of state. Thus  the gravitational term has the same sign as in the spherically symmetric case. However its effect  is the inverse of the latter case. Indeed, since the p.g.m.d is negative, then the gravitational term tends to increase $D_TU$, i.e. it acts as a repulsive force instead of  an attractive one, as in (C6) of \cite{epjc}. In the same order of ideas we see that a negative pressure gradient would tend to push any fluid element inwardly, i.e. everything happens as if force terms switch their roles, as compared with the spherically symmetric case.

\section{THE TRANSPORT EQUATION}
The treatment of dissipative processes requires the adoption of a heat transport equation. In order to ensure causality we shall resort to   the  transport equation obtained form 
the  M\"{u}ller--Israel--Stewart  theory\cite{Is,Is2,Is3}.

Then, the  corresponding  transport equation for the heat flux reads

\begin{equation}
\tau
h^{\alpha\beta}V^{\gamma}q_{\beta;\gamma}+q^{\alpha}=-\kappa h^{\alpha\beta}
(T_{,\beta}+Ta_{\beta}) -\frac 12\kappa T^2\left( \frac{\tau
V^\beta }{\kappa T^2}\right) _{;\beta }q^\alpha ,  \label{21t}
\end{equation}
where $\kappa $  denotes the thermal conductivity, and  $T$ and
$\tau$ denote temperature and relaxation time respectively. 

There is only one non-vanishing independent component of  Equation (\ref{21t}),  which may be written as
\begin{equation}\label{ecTra}
  \tau D_Tq=-q-\frac{\kappa}{AB}(AT)^\prime-\frac{1}{2}\tau \Theta q-\frac{1}{2}\kappa T^2D_T\left(\frac{\tau}{\kappa T^2}\right)q.
\end{equation}
In the case $\tau=0$ we recover the Eckart--Landau equation \cite{17T}.

Under some circumstances it is possible to adopt  the so called ``truncated'' version where the last term in (\ref{21t}) is neglected \cite{PAN},
\begin{equation}
\tau
h^{\alpha\beta}V^{\gamma}q_{\beta;\gamma}+q^{\alpha}=-\kappa h^{\alpha\beta}
(T_{,\beta}+Ta_{\beta}) \label{V1},
\end{equation}
and whose    only non--vanishing independent component becomes
\begin{equation}
\tau \dot q+qA=-\frac{\kappa}{B}(TA)^{\prime}. \label{V2}
\end{equation}

Let us now analyze in some detail the changes appearing in  the condition for thermal equilibrium, as compared with the spherically symmetric case.

As it was pointed out by Tolman many years ago \cite{Tol}, the fact established by  special relativity that all forms
of energy have inertia, should also apply to heat. Then the
equivalence principle implies that there should  be also some weight associated to heat, and one
should expect that thermal energy tends to displace to regions of lower gravitational
potential. Therefore the condition of thermal equilibrium in the
presence of a gravitational field must change with respect to its form in the absence
of gravity. More specifically, a temperature gradient is necessary in thermal equilibrium in order
to prevent the flow of heat from regions of higher to lower gravitational potential.

Thus Tolman condition reads
\begin{equation}
\left(TA\right)^\prime=0 \Rightarrow T^\prime=-\frac{T}{A}A^\prime=-TaB.
\label{t1}
\end{equation}
However as it follows from (\ref{EE11a}), if $m>4\pi P_r R^3$, in equilibrium $a<0$, (the four--acceleration  is now directed radially inwardly), implying the existence of a repulsive   gravitational force, leading to a positive temperature gradient in order to assure thermal equilibrium. This situation is at variance with the spherically symmetric case, where a negative temperature gradient is required to assure thermal equilibrium.

Before concluding this section it is worth  discussing about the physical implications of (\ref{pprimaU3}). This equation comes out from the combination of the dynamical equation (\ref{pprimaU2}) and the transport equation. It brings out the thermal effect on the p.g.m.d., and by virtue of the equivalence principle, on the effective inertial mass density as well. A similar effect was pointed out for the first time for the spherically symmetric case in \cite{1tef} (see also \cite{ther})  for a discussion on this effect). In our case the term $\frac{\kappa T}{\tau}$ increases the absolute value of the effective p.g.m.d (which is negative), thereby increasing the absolute value of the effective inertial mass density (the term in the bracket on the left of (\ref{pprimaU3})), as a result of which  any hydrodynamic force directed outward tends to push  the fluid element  inward, weaker  than in the non--dissipative case, due to the term $\frac{\kappa T}{\tau}$. On the other hand the gravitational term which is negative push any fluid element as it does in the non--dissipative case. Overall, the thermal effect enhance the tendency to expansion   as in the spherically symmetric case, but different terms in the equation playing different roles as compared with  this latter case.

In order to obtain specific solutions to the Einstein equations we shall need to impose additional restrictions. In this work we shall assume that the fluid evolves in the quasi--homologous regime and satisfies the vanishing complexity factor condition. The next two sections are devoted to explain these conditions in some detail.

\section{The structure scalars and the complexity factor}
The complexity factor is a scalar function intended to measure the degree of complexity of a self-gravitating system  (in some cases more than one scalar function may be required).
For a static, hyperbolically symmetric fluid distribution it was assumed in \cite{st1} (following the arguments developed in \cite{c1}) that the simplest system corresponds  to a 
 homogeneous (in the energy density), locally  isotropic fluid distribution (principal stresses equal). Thus, a zero value of  the
 complexity factor was assumed for such a distribution. Furthermore, it was shown that   a single scalar function (hereafter referred to as $Y_{TF}$) describes the modifications introduced by the energy density inhomogeneity and pressure anisotropy,   to the Tolman mass, with respect to its value  for the zero complexity case.

This  scalar belongs to   a set of variables named  structure scalars and defined in \cite{20}, and which appear in the orthogonal splitting of the Riemann tensor \cite{18, 18b, 18c, 19}.  For the sake of completeness we shall highlight the main steps leading to their acquisition (for the spherically symmetric case see \cite{20, 2cd} for details). For our purpose here, we shall need only one of the five structure scalars characterizing   our fluid distribution.

The first step consists in defining the tensor $Y_{\alpha\beta} $ by

\begin{eqnarray}
Y_{\alpha\beta} = R_{\alpha \gamma \beta \delta}V^\gamma
V^\delta,\label{Telectric} \end{eqnarray}

which may be splitted in terms of its trace and its trace free part as

\begin{eqnarray}
Y_{\alpha\beta} = \frac{1}{3}Y_T h_{\alpha
\beta}+Y_{TF}\left(\chi_{\alpha} \chi_{\beta}-\frac{1}{3}h_{\alpha
\beta}\right).\label{Telectricb}
\end{eqnarray}

Then using the field equations and (\ref{escE}) the following expressions can be obtained
\begin{eqnarray}
Y_T=4\pi(\mu+3 P_r-2\Pi) , \qquad
Y_{TF}={\cal E}-4\pi \Pi .\label{TEY}
\end{eqnarray}
On the other hand, combining   (\ref{EE11}),(\ref{EE22}),(\ref{escE}) and  (\ref{fmasa})  we obtain
\begin{equation}\label{3m/R32}
  \frac{3m}{R^3}=-4\pi \mu+4\pi \Pi+\mathcal{E},
\end{equation}
or using  (\ref{3m/R3}) and (\ref{TEY})
\begin{equation}\label{YTF}
  Y_{TF}=-8\pi \Pi+\frac{4\pi}{R^3} \int^r_0{R^3\left(D_R{ \mu}-3 q \frac{U}{RE}\right) R^\prime dr}.
\end{equation}
Using (\ref{EE11}), (\ref{EE22}) and  (\ref{escE}), we may express $Y_{TF}$ in terms of the metric functions and their derivatives

\begin{eqnarray}
  Y_{TF}&=&\frac{1}{B^2}\left (\frac{A^{\prime\prime}}{A}-\frac{A^\prime}{A}\frac{R^\prime}{R}-\frac{A^\prime}{A}\frac{B^\prime}{B}\right)\nonumber\\
 &+& \frac{1}{A^2}\left (\frac{\dot A}{A}\frac{\dot B}{B}- \frac{\dot A}{A}\frac{\dot R}{R} -\frac{\ddot{B}}{B}+\frac{\ddot{R}}{R}\right ).\label{ecYTF}
\end{eqnarray}

Following the arguments presented in \cite{st1} we shall choose $Y_{TF}$ as the complexity factor.
In the dynamic case, however, we still need to provide a criterion for the definition of complexity  of the pattern of evolution.

 We shall assume here that $Y_{TF}$  is identified with the complexity factor, and we shall consider the
 quasi-homologous evolution defined in \cite{epjc} as the simplest mode of evolution.

\section{The quasi--homologous condition}
In order to provide a rigorous  definition of  quasi--homologous evolution, let us write (\ref{EE01}) as
 \begin{equation}
 \left(\frac{U}{R}\right)^\prime=4\pi q B+\sigma\frac{R^\prime}{R},
 \label{qsh1}
 \end{equation}
whose general solution is

 \begin{equation}
U=\tilde a(t) R+R\int^r_0{\left (\frac{4\pi q }{E}+\frac{\sigma}{R}\right)R^\prime}dr,
 \label{qsh2}
 \end{equation}
where $\tilde a(t)$ is an integration function and (\ref{20x}) has been used.

Assuming that our fluid distribution is bounded by a surface $\Sigma^e$ defined by the equation $r=r_{\Sigma^e}=constant$, we may write

 \begin{equation}
U=R\frac{U_{\Sigma^e}}{R_{\Sigma^e}}-R\int^{r_{\Sigma^e}}_r{\left (\frac{4\pi q }{E}+\frac{\sigma}{R}\right)R^\prime}dr.
 \label{qsh3}
 \end{equation}

The quasi--homologous condition  reads
\begin{equation}
U=R\frac{U_{\Sigma^e}}{R_{\Sigma^e}},
 \label{qsh4bis}
 \end{equation}
implying

 \begin{equation}
\frac{4\pi q }{E}+\frac{\sigma}{R}=0.
 \label{qsh4}
 \end{equation}
 
The above condition will be used to obtain specific models, and its assumption is supported, on the one hand, by the fact that it is the relativistic version of the well-known homologous condition widely used in classic astrophysics, and on the other hand by the fact that it qualifies as   one of the simplest  patterns of evolution (see \cite{2cd, epjc}  for a discussion on this point).

\section{The exterior spacetime and junction conditions}
In the case that the fluid is bounded then junction conditions on the boundary have to be imposed \cite{Darmois} in order to avoid the presence of thin shells on the boundary.  If any specific  model does not satisfy the Darmois conditions  then we should relax  the continuity of the second fundamental form, which  would imply the presence of thin shells \cite{17}.

Thus, outside $\Sigma^e$ (but  inside the horizon) we assume that we have the hyperbolic version of the Vaidya
spacetime, described by
\begin{equation}
ds^2=-\left[\frac{2M(v)}{\bf r}-1\right]dv^2-2 d{\bf r} dv+{\bf r}^2(d\theta^2
+\sinh^2\theta
d\phi^2) \label{1int},
\end{equation}
where $M(v)$  denotes the total mass,
and  $v$ is the retarded time.

The continuity of the first fundamental form reads
\begin{equation}
(ds^2)^-_{\Sigma^e} = (ds^2)^+_{\Sigma^e},
\label{1ff}
\end{equation}
where $-$, $+$ meaning  from the inner or the outer side of the boundary surface respectively.

At the outer side of the boundary, the surface equation reads
\begin{equation}
\Psi\equiv {\bf r}-{\bf r}_{\Sigma^e}(v)=0,
\label{suro}
\end{equation}
whose unit normal vector is defined by

\begin{equation}
n^+_\mu=\frac{\partial_\mu \Psi}{\sqrt{\vert \partial_\alpha \Psi \partial _\beta\Psi g^{\alpha \beta}\vert}},
\label{no1}
\end{equation}
with components
\begin{equation}
n^+_\mu=\left(-\beta \frac{d{\bf r}_{\Sigma^e}}{dv}, \beta, 0, 0 \right),
\label{no2}
\end{equation}
where
\begin{equation}
\beta=\frac{1}{\sqrt{\vert \frac{2M(v)}{{\bf r}_{\Sigma^e}}-1+\frac{2d{\bf r}_{\Sigma^e}}{dv}\vert}}.
\label{no3}
\end{equation}

At the inner side, the surface equation reads

\begin{equation}
\Phi\equiv r-r_{\Sigma^i}=0,
\label{suro}
\end{equation}
whose normal unit vector is defined by
\begin{equation}
n^-_\mu=\frac{\partial_\mu \Phi}{\sqrt{\vert \partial_\alpha \Phi \partial _\beta\Phi g^{\alpha \beta}\vert}},
\label{ni1}
\end{equation}
with components
\begin{equation}
n^-_\mu=\left(0, B_\Sigma, 0, 0 \right),
\label{ni2}
\end{equation}
observe that $n^-_\mu=(\chi_\mu)_\Sigma$.

From (\ref{1ff}) it follows that
\begin{equation}
R(t, r_{\Sigma^e})={\bf r}_{\Sigma^e}(v).
\label{ifffb}
\end{equation}

Next, instead of calculating the second fundamental form at both sides of the boundary surface we shall impose the continuity of the flux of energy--momentum across $\Sigma^e$, which of course implies the absence of thin shells on the boundary surface. For doing so we have to calculate
\begin{eqnarray}
(T_{\mu \nu} n^\nu n^\mu)^+_{\Sigma^e}; \quad (T_{\mu \nu} n^\nu n^\mu)^-_{\Sigma^e}; \nonumber \\(T_{\mu \nu} n^\nu V^\mu)^+_{\Sigma^e};\quad(T_{\mu \nu} n^\nu V^\mu)^-_{\Sigma^e}.
\label{flux}
\end{eqnarray}
where the vectors $(V^\mu)^+$ , $(V^\mu)^-$ have components
\begin{equation}
(V^\mu)^+=\left[\beta,\beta \frac{d{\bf r}_{\Sigma^e}(v)}{dv}, 0, 0\right],
\label{vext}
\end{equation}
and
\begin{equation}
(V^\mu)^-=\left[\frac{1}{A},0, 0, 0\right].
\label{vext}
\end{equation}

Next, we have to calculate the energy--momentum tensor corresponding to the line element  (\ref{1int}), we obtain
\begin{equation}\label{Tint}
  T_{\mu\nu}^{(+)}=\frac{1}{4\pi {\bf r}^2}\frac{dM}{dv}\delta ^0_\mu \delta ^0_\nu.
\end{equation}

From the above expression it follows at once that the energy density of the null fluid sourcing (\ref{1int}) would be negative for an outgoing flux, which is exactly the inverse of what happens for the usual Vaidya metric.

We can now evaluate (\ref{flux}) to obtain
\begin{eqnarray}
  (T_{\mu\nu}n^\mu n^\nu)^{-}_{\Sigma^e} &=& [P_r]_{\Sigma^e}, \\
   (T_{\mu\nu}n^\mu V^\nu)^{-}_{\Sigma^e} &=&- [q]_{\Sigma^e}, \\
   (T_{\mu\nu}n^\mu n^\nu)^{+}_{\Sigma^e} &=& \frac{\beta^2}{4\pi {\bf r}^2_{\Sigma^e}}\frac{dM}{dv}, \\
   (T_{\mu\nu}n^\mu V^\nu)^{+}_{\Sigma^e} &=& -\frac{\beta^2}{4\pi {\bf r}^2_{\Sigma^e}}\frac{dM}{dv}.
\end{eqnarray}

Then imposing the continuity of the flux of energy--momentum across $\Sigma^e$, it follows that
\begin{equation}
q\stackrel{\Sigma^e}{=}P_r.\label{j3}
\end{equation}
where $\stackrel{\Sigma^e}{=}$ means that both sides of the equation
are evaluated on $\Sigma^e$.

Finally, following the usual procedure used in the spherically symmetric case, it is a simple matter to check that the continuity of the second fundamental form implies
 \begin{equation}
m(t,r)\stackrel{\Sigma^e}{=}M(v). \label{junction1}
\end{equation}

In the cases where the central region is surrounded by an empty vacuole bounded by a surface $\Sigma^i$, junction conditions should be considered also at the inner boundary of the fluid distribution. Then, following the same steps as before we find
\begin{equation}
P_r\stackrel{\Sigma^i}{=}0.\label{j3b}
\end{equation}
and
 \begin{equation}
m(t,r)\stackrel{\Sigma^i}{=}0. \label{junction1b}
\end{equation}
\

\section{Some models}

In the following subsections we shall exhibit several families of solutions to the Einstein equations for hyperbolically symmetric fluids. These solutions will be obtained  by assuming quasi--homologous evolution and the vanishing of the complexity factor. This choice, justified by the comments in previous sections, will allow us to compare the behavior of our fluid distributions with  similar solutions found for the spherically symmetric case in \cite{epjc}. Besides, different types of additional restrictions will be imposed in order to obtain specific models. It should clear that the purpose of the presentation of these models, besides the possible potential of some of them in the study of specific astrophysical scenario, is to illustrate the richness of fluid distributions endowed with hyperbolical symmetry.

\subsection{ Non--dissipative case}
Excluding dissipative processes, and assuming the quasi--homologous condition (\ref{qsh4}) we may write

\begin{equation}\label{qcero}
  q=0\,\,\Rightarrow\,\, \sigma=0\,\,\Rightarrow\,\, \frac{\dot B}{B}=\frac{\dot R}{R}\,\,\Rightarrow\,\, R=rB,
\end{equation}
and using (\ref{qsh2})

\begin{equation}\label{chomo}
  U=\frac{\dot R}{A}=\frac{r\dot B}{A}={\tilde a}(t)rB.
\end{equation}
Imposing next the condition  $Y_{TF}=0$ we have

\begin{equation}\label{YTFcero1}
  \frac{A^{\prime\prime}}{A}-\frac{A^\prime}{A}\frac{B^\prime}{B}-\frac{A^\prime}{A}\frac{R^\prime}{R}=0.
\end{equation}

In order to exhibit specific  solutions, we shall further assume some additional restrictions.

\subsubsection{ $\mathcal{E}=0$, $\Pi=0$ }
We shall assume here that the fluid is conformally flat ($\mathcal{E}=0$) and the pressure is isotropic ($\Pi=0$), which combined with $Y_{TF}=0$ produces $\mu^\prime=0$ (i.e. the energy density is homogeneous).

From the  conditions   $\mathcal{E}=0$ and   $\Pi=0$ we obtain
\begin{equation}\label{ecm1}
  \frac{1}{R^2}+\frac{1}{B^2}\left[\left(\frac{R^\prime}{R}\right)^2+\frac{B^\prime}{B}\frac{R^\prime}{R}-\frac{R^{\prime\prime}}{R} \right]-\frac{1}{A^2}\left( \frac{\dot R^2}{R^2}-\frac{\dot B}{B}\frac{\dot R}{R}
  \right)=0.
\end{equation}

Using  (\ref{qcero}) in  (\ref{ecm1}) and  (\ref{YTFcero1}) produces
\begin{eqnarray}
 1+r^2\left[ 2\left(\frac{R^\prime}{R}\right)^2-\frac{1}{r}\frac{R^\prime}{R}-\frac{R^{\prime\prime}}{R} \right ]= 0\label{secm1},
 \end{eqnarray}
 and
  \begin{eqnarray}
  \frac{A^{\prime \prime}}{A}-\frac{A^\prime}{A}\left ( \frac{2R^\prime}{R}-\frac{1}{r}\right ) = 0.\label{secm2}
\end{eqnarray}
The solution to the system  (\ref{secm1})--(\ref{secm2}) is easily found to be
\begin{eqnarray}\label{nds1}
  R &=& \frac{\tilde R(t)}{\cos[c_1(t)+\ln r]}, \\
  B &=&  \frac{\tilde R(t)}{r \cos[c_1(t)+\ln r]},  \\
  A &=&\gamma(t) \tilde R^2(t) \tan [c_1(t)+\ln r]+b(t), \label{nds7}
\end{eqnarray}
where $\tilde R(t), c_1(t), \gamma(t), b(t)$ are arbitrary functions of their argument. The reader can easily check, using Maple or Mathematica, that the line element (\ref{1}) with (\ref{nds1})--(\ref{nds7}) produces   $\mathcal{E}=0=\Pi=0$.

To specify further the  solution we shall  choose the above functions as follows

\begin{equation}\label{condFRW}
  \dot c_1=\frac{\dot{\tilde R}}{\tilde R}, \qquad b(t)=\gamma(t)\tilde R^2,
\end{equation}
producing

\begin{eqnarray}
  \frac{\dot R}{R} &=& \frac{\dot{\tilde R}}{\tilde R}(1+\tan u), \\
  A &=& \gamma(t)\tilde R^2(1+\tan u),\quad \Rightarrow \quad A= \frac{\tilde a\dot R}{R}\label{condA},
\end{eqnarray}
with  $\tilde a=\frac{\gamma(t)\tilde R^3}{\dot{\tilde R}}$ and  $u=c_1(t)+\ln r$.
From  the above expressions we found for the physical variables and the mass function,

\begin{eqnarray}
  8\pi \mu &=& -\frac{3}{\tilde R^2}+\frac{3}{\tilde a^2}, \\
  8\pi P_r =8\pi P_\bot&=&-\frac{3}{\tilde a^2}+\frac{3\tan u+1}{\tilde R^2(\tan u+1)}\nonumber 
  \\ &+&\frac{2\tilde R \dot{\tilde a}}{\tilde a^3 \dot{\tilde R}(\tan u+1)},\label{nds2}
  \\
  m &=& \frac{\tilde{R}}{2 \cos ^3 u}\left(1-\frac{\tilde {R}^2}{\tilde{a}^2}\right).
\end{eqnarray}

It a simple matter to check that  this solution does not satisfy the Darmois conditions at either boundary surfaces and therefore we must assume the presence of thin shells there.

If we choose  $\tilde R(t), c_1(t), \gamma(t)$ such that they tend to a constant as $t\rightarrow \infty$, then the above solution tend to the incompressible isotropic solution found in \cite{mimII}, which is a particular case of the hyperbolically symmetric Bowers--Liang solution found in \cite{st1}.

The above solution might be considered as a version of the Friedman--Robertson--Walker space--time (FRW)  for the hyperbolically symmetric case since they  share some  similar properties e.g. $\mathcal{E}=\Pi=\mu^\prime=\sigma=0$. However it is not geodesic as in the spherically symmetric case. Therefore we shall next find another  version of the hyperbolically symmetric  FRW space--time, but satisfying the geodesic condition $A^\prime=0$.
\subsubsection{Geodesic solutions}

If we further impose the geodesic condition on the fluid, then we may put  without loss of generality $A=1$ and the quasi--homologous condition also implies
\begin{equation}
\frac{R_I}{R_{II}} = constant,
\label{g1}
\end{equation}
where $R_I$ and $R_{II}$ denote the areal radii of two shells $(I, I I)$ described by $r = r_I = constant$, and $r = r_{II} = constant$, respectively. 

From (\ref{g1}) it follows at once that  $R$ is a separable function. In the notation of \cite{2cd},  conditions (\ref{chomo}) and (\ref{g1}) define the homologous evolution.

The conditions $A=1$ and  $q=0$  imply

\begin{equation}\label{ecA1q0}
\frac{\dot{B}}{B}=\frac{\dot{R}^\prime}{R^\prime},
\end{equation}
where (\ref{EE01}) has been used.
Since the fluid is shear--free we have $R=Br$, and since $R$ is separable so is $B$. But if $B$ is separable, then by a simple reparametrization of $r$ it becomes a function of $t$ alone $B=B(t)$, i.e.
\begin{equation}\label{YTF0A1}
 R=rB(t).
\end{equation}
Then (\ref{ecA1q0}) is automatically satisfied,  as well as  $Y_{TF}=0$ as it follows from  (\ref{YTFcero1}).
In this case we may write the physical variables and the mass function  as
\begin{eqnarray}
8\pi\mu&=&-\frac{2}{r^2 B^2}+\frac{3 \dot{B}^2}{B^2},
\label{1b}\\
8\pi P_r&=&\frac{2}{r^2 B^2}-\frac{\dot{B}^2}{B^2}-\frac{2\ddot{B}}{B},
\label{2b}\\
8\pi P_{\bot}&=&-\frac{2\dot{B}^2}{B^2}-\frac{2\ddot{B}}{B},
\label{3b}\\
m&=&\frac{rB}{2}(2-r^2 \dot {B}^2).
\label{4b}
\end{eqnarray}
Thus the fluid  is conformally flat, shear--free, geodesic, evolves homologously and satisfies the  vanishing complexity factor condition. In this sense it could be considered also as a version of the hyperbolically symmetric FRW space--time. However, unlike the spherically symmetric case, it is anisotropic and the energy--density is inhomogeneous.

As in the previous solution, by simple inspection of (\ref{2b}), (\ref{4b})  it can be checked that Darmois conditions cannot be satisfied at either boundary surface.

It is worth analyzing with some detail the differences between this case and the situation in  the spherically  symmetric case (the usual one). In the latter case we have seen \cite{2cd} that for a non--dissipative fluid satisfying the homologous condition, the complexity factor vanishes and there is a single solution characterized by $\Pi=\mu^\prime=a={\cal E}=0$ (FRW).

However in the present case, imposing homologous condition on a geodesic non--dissipative fluid we get a conformally flat, shear--free geodesic fluid  with  $\Pi, \mu^\prime\neq 0$.
If we want to describe an isotropic, homogeneous, shear--free non--dissipative  fluid, then we have to relax the geodesic condition.

Finally, it is instructive to build up a toy model with the above solution, by choosing a particular form for the function $B$ such that asymptotically it leads to a static regime. Thus, let us assume.
\begin{equation}
B=\beta\left(1+e^{-\alpha t}\right),
\label{t1}
\end{equation}
where $\alpha, \beta$ are two positive constants.

Then it  is a simple matter to check that as $t\rightarrow \infty$ we get
\begin{eqnarray}
  8\pi \mu &=& -\frac{2}{r^2 \beta^2},
  \\
  8\pi P_r &=& \frac{2}{r^2 \beta^2},\\
  8\pi P_\bot &=&0,
\end{eqnarray}
and for the mass function we get asymptotically $m=r\beta$.

Thus our toy model converges to the static solution corresponding to the stiff equation of state $(P_r=\vert \mu \vert)$ found in \cite{st1} (Eqs.(138-139) in that reference).

We shall next consider dissipative solutions.

\subsection{Dissipative case with  $B=1$}
Let us now consider  dissipative solutions satisfying the condition $B=1$. As discussed in \cite{16n}, such a condition is particularly suitable for describing fluid distributions whose center is surrounded by an empty cavity, a scenario we expect for the kind of fluid distributions we are dealing with in this work.

Thus, the metric functions for this case read
\begin{equation}
B=1,\qquad A=\frac{\dot R}{\tilde a(t) R},
\label{mis1}
\end{equation}
and the corresponding Einstein equations may be written as

\begin{widetext}
\begin{eqnarray}\label{EEB1}
 8\pi  \mu &=& -\frac{1}{R^2}-\frac{2 R^{\prime\prime}}{R}-\left(\frac{R^\prime}{R}\right)^2+\tilde a^2, \\
  4\pi q &=& \frac{\tilde a(t){R}^\prime}{R}, \label{101}\\
  8\pi  P_r &=& \frac{1}{R^2} -\left(\frac{R^\prime}{R}\right)^2+\frac{2\dot R^\prime R^\prime}{\dot R R} -3\tilde a^2-2\dot {\tilde a} {\tilde a} \frac{R}{\dot R},\\
  8\pi P_\bot &=&  \frac{\dot R^{\prime\prime}}{\dot R}-\frac{\dot{R^\prime}}{\dot R}\frac{R^\prime}{R}+\left(\frac{R^\prime}{R}\right)-\tilde a^2-\dot {\tilde a} {\tilde a} \frac{R}{\dot R}. 
\end{eqnarray}
\end{widetext}
We may formally integrate (\ref{V2}) producing for the temperature
\begin{equation}
T(t,r)=\frac{\tilde a R}{\dot R}\left(f(t) -\frac{\tau \dot{\tilde a}}{4\pi \kappa}\ln R-\frac{1}{4\pi \kappa} \int{\frac{\dot R}{R}\frac{R^\prime}{R}dr}\right)-\frac{\tau \tilde a^2}{4\pi \kappa},
\label{mis2}
\end{equation}
where $f(t)$ is a function of integration.
On the other hand the condition $Y_{TF}=0$  and (\ref{qsh4}), now read
\begin{eqnarray}
  A^{\prime\prime}-A^\prime \frac{R^\prime}{R}+A\sigma ^2&=&\dot{\sigma}, \label{cond1}\\
   -\frac{\dot{R}}{\sigma R}&=&A.\label{cond2}
\end{eqnarray}

Introducing the intermediate variables $(X, Y)$,
\begin{equation}\label{ccam}
  A=X+\frac{\dot{\sigma}}{\sigma^2}\quad  {\rm and} \quad R=X^\prime Y,
\end{equation}
(\ref{cond1})  and (\ref{cond2}) become

\begin{equation}\label{Cond1P}
  -\frac{X^\prime}{X}\frac{Y^\prime}{Y}+\sigma^2=0,
\end{equation}

\begin{equation}\label{Cond2P}
  \frac{\dot X^\prime}{X^\prime}+\frac{\dot Y}{Y}=-\sigma X-\frac{\dot{\sigma}}{\sigma}.
\end{equation}

Thus we have a large family of dissipative solutions, among which we shall select some specific ones, by imposing additional restrictions allowing us to integrate (\ref{Cond1P}) and (\ref{Cond2P}).
\subsubsection{$X$ is a separable function}
If we assume the  function $X$ to be separable, then we can integrate the system (\ref{Cond1P}) and (\ref{Cond2P}), obtaining
 \begin{eqnarray}
  A&=&\frac{\dot{\sigma}}{2\beta^2\sigma^2}\left [ 2\beta^2-\sigma^2(\beta r+c_1)^2\right],  \\
   R&=&\frac{\tilde{R}_0}{\sigma}(\beta r+c_1)e^{\frac{\sigma^2 }{4\beta^2}(\beta r+c_1)^2}\label{RB1},\\
   \tilde{a}&=&-\sigma,
\end{eqnarray}
where $\beta$,  $\tilde{R_0}$ and $c_1$ are constants.

The above expressions allow us to write for the physical variables

\begin{equation}\label{EEB1}
 8\pi  \mu = -\frac{\sigma^2 e^{-\frac{\sigma^2}{2\beta ^2}(\beta r+c_1)^2}}{\tilde{R}_0^2(\beta r+c_1)^2}-\frac{\beta^2}{(\beta r+c_1)^2}-\frac{3\sigma^4}{4\beta^2}(\beta r+c_1)^2-3\sigma^2,
 \end{equation}
\begin{eqnarray}\label{EEB2}
  4\pi q = -\frac{\sigma[2\beta^2+\sigma^2 (\beta r+c_1)^2]}{2\beta(\beta r+c_1)},
 \end{eqnarray}
\\ 
  \begin{eqnarray}\label{EEB3}
  8\pi  P_r& =& \frac{\sigma^2 e^{-\frac{\sigma^2}{2\beta ^2}(\beta r+c_1)^2}}{\tilde{R}_0^2(\beta r+c_1)^2}-\frac{4\sigma^2\beta^2}{2\beta^2-\sigma^2(\beta r+c_1)^2}+\frac{\beta^2}{(\beta r+c_1)^2}\nonumber \\&+&\frac{\sigma^4}{4\beta^2}(\beta r+c_1)^2,
  \end{eqnarray}
\begin{equation}\label{EEB4}
  8\pi P_\bot =-\frac{\sigma^2[2\beta^2+\sigma^2(\beta r+c_1)^2]^2}{4\beta^2(2\beta^2-\sigma^2(\beta r+c_1)^2)},
\end{equation}
\begin{widetext}
\begin{equation}\label{EEmas}
  m=\frac{\tilde{R}_0(\beta r+c_1)}{2\sigma}e^{\frac{\sigma^2}{4\beta ^2}(\beta r+c_1)^2}
  \left \{1+\frac{\tilde{R}^2_0}{4\sigma^2\beta^2}\left[4\beta^4+\sigma^4 (\beta r+c_1)^4\right]e^{\frac{\sigma^2}{2\beta ^2}(\beta r+c_1)^2}   \right \},
\end{equation}
while the expression for the temperature reads in this case as

\begin{eqnarray}
  T(t,r)&=&\frac{2\beta^2\sigma^2}{\dot{\sigma}\left[2\beta^2-\sigma^2(\beta r+c_1)^2\right]}\left\{f(t)+\frac{\dot{\sigma}
   \tau}{4\pi \kappa}\left[\frac{\sigma^2 }{4\beta^2}(\beta r+c_1)^2+\ln \left[\frac{\tilde{R}_0}{\sigma}(\beta r+c_1)\right]\right]
 \right .\nonumber
  \\
 &+&\left .   \frac{\dot{\sigma}}{4\pi \sigma \kappa}\ln(\beta r+c_1)
 -\frac{\dot{\sigma}\sigma^3}{64\pi \beta^4\kappa}(\beta r+c_1)^4\right \}-\frac{\tau \sigma^2}{4\pi  \kappa}.\label{TB1M1}
\end{eqnarray}
\end{widetext}

\subsubsection{$A=A(r)$}
Another sub--family of solutions may be obtained  by assuming that $A$ only depends on $r$, then the solution to the system (\ref{Cond1P}) and (\ref{Cond2P}) produces
 \begin{eqnarray}
  A&=&\frac{1}{4}(\sqrt{2\sigma_0}r+c_1)^2, \qquad \tilde{a}=\sigma_0 t-\sigma_1,\\
   R&=&\tilde{R}(r)e^{-\frac{1}{4}(\sqrt{2\sigma_0}r+c_1)^2(-\frac{\sigma _0}{2}t^2+\sigma_1 t)},\label{RB2}
\end{eqnarray}
where $\tilde R(r)$ is an arbitrary function of its argument, and $\sigma_0, \sigma_1, c_1$  are constants.
To obtain a specific model, we shall further assume $\tilde R=\tilde R_0 ={\rm constant}$, in which case  we find for the physical variables
\begin{widetext}
\begin{eqnarray}\label{EEB2}
 8\pi  \mu =\sigma_1^2-\frac{3\sigma_0}{2}(\sqrt{2\sigma_0}r+c_1)^2\left(-\frac{\sigma_0}{2}t^2+\sigma_1 t\right)^2-\frac{1}{\tilde{R}_0^2}e^{\frac{1}{2}(\sqrt{2\sigma_0}r+c_1)^2(-\frac{\sigma _0}{2}t^2+\sigma_1 t)},
 \end{eqnarray}
\begin{eqnarray}\label{EEB3}
  4\pi q =\frac{\sqrt{2\sigma_0}}{2}(\sqrt{2\sigma_0}r+c_1)\left(-\frac{\sigma_0}{2}t^2+\sigma_1 t\right)(-\sigma_0 t+\sigma_1),
\end{eqnarray}
 \begin{eqnarray}\label{EEB4}
  8\pi  P_r &=&\frac{1}{\tilde{R}_0^2}e^{\frac{1}{2}(\sqrt{2\sigma_0}r+c_1)^2(-\frac{\sigma _0}{2}t^2+\sigma_1 t)}-t^2\sigma_0^2+2t\sigma_0\sigma_1-3\sigma_1^2
  -\frac{8\sigma_0}{(\sqrt{2\sigma_0}r+c_1)^2}\nonumber \\&+&\frac{\sigma_0}{2}(\sqrt{2\sigma_0}r+c_1)^2\left(-\frac{\sigma_0}{2}t^2+\sigma_1 t\right)^2,
\end{eqnarray}

\begin{equation}\label{EEB5}
  8\pi P_\bot =\frac{1}{2}\sigma_0^2 t^2-t\sigma_0\sigma_1-\sigma_1^2 +\frac{\sigma_0}{2}(\sqrt{2\sigma_0}r+c_1)^2\left(-\frac{\sigma_0}{2}t^2+\sigma_1 t\right)^2,
\end{equation}

\begin{equation}\label{EEmas5}
  m=\frac{\tilde{R}_0}{2}e^{-\frac{1}{4}(\sqrt{2\sigma_0}r+c_1)^2(-\frac{\sigma_0}{2}t^2+\sigma_1 t)}
  \left\{1+\tilde{R}_0^2\left[\frac{\sigma_0}{2}(\sqrt{2\sigma_0}r+c_1)^2(-\frac{\sigma_0}{2}t^2+\sigma_1 t)^2-(\sigma_0 t-\sigma_1)^2\right]
  e^{-\frac{1}{2}(\sqrt{2\sigma_0}r+c_1)^2(-\frac{\sigma_0}{2}t^2+\sigma_1 t)}\right \}.
\end{equation}
For the temperature the corresponding expression reads

\begin{eqnarray}
  T(t,r)&=&\frac{4}{(\sqrt{2\sigma_0}r+c_1)^2}\left \{f(t)
  -\frac{\tau \sigma_0}{4\pi \kappa}\left[\ln \tilde{R}_0-\frac{1}{4}(-\frac{\sigma_0}{2}t^2+\sigma_1 t)(\sqrt{2\sigma_0}r+c_1)^2\right]\right \}\nonumber
  \\
 &-&\frac{(-\frac{\sigma_0}{2}t^2+\sigma_1 t)(-\sigma_0 t+\sigma_1)(\sqrt{2\sigma_0}r+c_1)^2}{32\pi\kappa }
 -\frac{\tau (-\sigma_0 t+\sigma_1)^2 }{4\pi \kappa }.\label{TB1M2}
\end{eqnarray}

\end{widetext}

\subsubsection{$\dot{\sigma}=0$}
Finally, we shall obtain a class of solutions by assuming that the shear scalar is constant, in which case the integration of the system (\ref{Cond1P}) and (\ref{Cond2P}) produces

 \begin{eqnarray}
  A&=&\beta r-\frac{\beta^2}{\sigma}t+\beta_0, \qquad \tilde{a}=-\sigma=const. \\
   R&=&\tilde{R}_0\beta e^{\left(\frac{\sigma^2}{2}r^2-\sigma\beta t r+\frac{\sigma ^2 \beta_0}{\beta}r+\frac{\beta^2}{2}t^2-\sigma\beta_0 t\right)},\label{RB3}
\end{eqnarray}
where $\tilde{R}_0, \beta, \beta_0$ are constants.
The physical variables for this case read

\begin{eqnarray}\label{EEB2}
 8\pi  \mu &=&-\sigma^2-3\left[\sigma^2\left(r+\frac{\beta_0}{\beta}\right)-\sigma \beta t\right]^2\nonumber \\&-&\frac{ e^{-2\left(\frac{\sigma^2}{2}r^2-\sigma\beta t r+\frac{\sigma ^2 \beta_0}{\beta}r+\frac{\beta^2}{2}t^2-\sigma\beta_0 t\right)}}{\tilde{R}^2_0 \beta^2},
 \end{eqnarray}
 \begin{equation}\label{EEB3}
  4\pi q =-\sigma^3\left(r-\frac{\beta}{\sigma}t+\frac{\beta_0}{\beta}\right),
  \end{equation}
\\
 \begin{eqnarray}\label{EEB4}
  8\pi  P_r& =&-\sigma^2+\sigma^4\left(r-\frac{\beta}{\sigma}t+\frac{\beta_0}{\beta}\right)^2\nonumber \\&+&\frac{ e^{-2\left(\frac{\sigma^2}{2}r^2-\sigma\beta t r+\frac{\sigma ^2 \beta_0}{\beta}r+\frac{\beta^2}{2}t^2-\sigma\beta_0 t\right)}}{\tilde{R}^2_0 \beta^2},
  \end{eqnarray}
   \begin{equation}\label{EEB5}
    8\pi P_\bot =\sigma^2+\left[\sigma^2\left(r+\frac{\beta_0}{\beta}\right)-\sigma \beta t\right]^2,
\end{equation}
and

\begin{widetext}

\begin{equation}\label{EEmas6}
  m=\frac{\tilde{R}_0\beta}{2}  e^{\left(\frac{\sigma^2}{2}r^2-\sigma\beta t r+
  \frac{\sigma ^2 \beta_0}{\beta}r+\frac{\beta^2}{2}t^2-\sigma\beta_0 t\right)}\left \{1+\tilde{R}_0^2\beta^2\sigma^2
  \left[\frac{\sigma^2}{\beta^2}(\beta r-\frac{\beta^2 t}{\sigma} + \beta_0)^2-1)\right]  e^{2\left(\frac{\sigma^2}{2}r^2-\sigma\beta t r+
  \frac{\sigma ^2 \beta_0}{\beta}r+\frac{\beta^2}{2}t^2-\sigma\beta_0 t\right)}  \right \},
\end{equation}

\begin{eqnarray}
  T(t,r)&=&\frac{f(t)}{ (\beta r-\frac{\beta^2}{\sigma} t+\beta_0)}
 +\frac{\sigma^3}{12\beta^2\pi\kappa}(\beta r-\frac{\beta^2}{\sigma} t+\beta_0)^2
 -\frac{\tau\sigma^2 }{4\pi \kappa}.\label{TB1M3}
\end{eqnarray}
\end{widetext}

It can be easily verified that none of the above solutions can be matched smoothly on either of boundary surfaces.
\section{Conclusions}
We have presented a general approach to describe the dynamics of hyperbolically symmetric fluids, including dissipative processes. Although our main motivation was (and still is) to provide a formalism allowing us to study the dynamic regime leading to a  static source of  the line element (\ref{w3}),  the obtained results are sufficiently general as to be applied  to any other scenario where we expect hyperbolical symmetry to play a relevant role.

The four  more remarkable features of hyperbolically symmetric fluids are:
\begin{enumerate}
\item The energy density is necessarily negative.
\item The fluid cannot fill the central region.
\item The Tolman condition for thermodynamic equilibrium implies in this case the presence of a positive temperature gradient. 
\item The  thermal modification of  the inertial mass density  reported for the spherically symmetric case in \cite{1tef}, produces an effect that is similar   to the one  obtained in the spherically symmetric case (to  enhance the tendency to expansion) but  comes about through  different terms in the equation.
\end{enumerate}

It should be reminded  that the first two  properties are common to the static and the dynamic regimes.

With respect to the violation of the weak energy condition ($\mu<0$) it should be stressed that  while it is true that at classical level we do not expect negative energy density in a realistic fluid, the situation is quite different at quantum regime, where the appearance of negative energy density  is possible  (see \cite{we1, we2, we3, cqg, pav} and references therein). This confirms our believe that the type of fluids considered in this manuscript  might be useful for studying systems under extreme conditions where quantum effects are expected to play a relevant role.

As  mentioned  in Section III,  this negative energy density implies the appearance of a repulsive  gravitational force which has two important thermodynamic consequences mentioned in the point 3 above.

Next, the impossibility of the fluid distribution to fill the central region leaves several possible scenarios. We  lean to assume the existence of an empty  vacuole  surrounding the center, however many other scenarios may be regarded as well, such as filling the central region with a fluid endowed with a different type of symmetry.
At any rate, this impossibility is consistent with the result obtained in \cite{2}, according to which test particles are not allowed to reach the center for the line element (\ref{w3}).

The final description of the central region, as well as the fulfillment or not of the Darmois conditions at both interfaces would depend on the specific system under consideration.

After having deployed the set of equations for describing the dynamics of hyperbolically symmetric fluids, we presented several exact solutions. These were found under the condition of the vanishing complexity factor defined in \cite{c1} ($Y_{TF}=0$) and the quasi--homologous evolution defined by (\ref{qsh4}).

We first considered the non--dissipative case. Two exact solutions were found for this case. One of them (\ref{nds1}--\ref{nds2}), describes a fluid distribution satisfying conditions $Y_{TF}={\cal E}=\sigma=0=\Pi=\mu^\prime=0$, which is a reminiscence of the usual FRW space--time. However, unlike the latter it is not geodesic. If we impose the geodesic condition,  then the quasi--homologous condition becomes homologous, and the solution is described by (\ref{YTF0A1})--(\ref{4b}). This is a geodesic fluid, satisfying also the conditions $Y_{TF}={\cal E}=\sigma=0$, and therefore is also a good candidate to be regarded as the hyperbolical version of the FRW space--time, however unlike the latter, it is anisotropic in the pressure and inhomogeneous in the energy--density. 

In both cases, if the arbitrary functions appearing in the solutions are chosen  such that the system tends to a static situation in the limit $t\rightarrow \infty$, then these solutions tend to the static solutions studied in \cite{st1}.

Thus alternative cosmological models emerge from the study of hyperbolically symmetric fluids, which could be of interest  when seeking for more sophisticated models of the Universe, (see for example \cite{apj} and references therein).

Finally, we considered the dissipative case. In order to obtain specific models we have restricted ourselves to the case where the condition $B=1$ is satisfied. Such a condition is suggested by the fact that it appears to be suitable for the description of fluids whose central region is surrounded by a vacuum cavity \cite{16n}. The purpose of these solutions, as well as the non--dissipative ones, is not the modeling of any  specific astrophysical scenario, but  just to illustrate a possible way of finding solutions, some of which might be used for the modeling of hyperbolically symmetric fluids required for describing specific physical situations. Neither of the exhibited models matches smoothly on  the boundary surfaces. In order to obtain models satisfying Darmois conditions, one could try to extend to the hyperbolically symmetric case, the general methods developed  for the spherically symmetric case in \cite{TM, TMb, ivanov1, ivanov}.

In the temperature profiles exhibited for each solution, we may identify two type of contributions. On the one hand  the contributions in the stationary dissipative regime (non containing $\tau$)  and the contributions from the transient regime (terms proportional to $\tau$).

\begin{acknowledgments}

This work was partially supported by Ministerio de
Ciencia, Innovaci\'on y Universidades. Grant number:
PGC2018–096038–B–I00, and Junta de Castilla y Le\'on.
Grant number: SA096P20.
\end{acknowledgments}

\appendix

\section{Conservation laws $T^\mu_{\nu;\mu}=0$}
In our case the conservation laws have only two independent components which read

\begin{widetext}
\begin{equation}\label{mupunto}
  \dot{\mu}+(\mu+P_r)\frac{\dot B}{B}+2(\mu+ P_\bot) \frac{\dot R}{R}+q^\prime \frac{A}{B}+2q\frac{A}{B}\left (
  \frac{A^\prime}{A}+\frac{R^\prime}{R}\right )=0,
\end{equation}
\end{widetext}
and

\begin{equation}\label{pprima}
  P_r^\prime+(\mu+P_r)\frac{A^\prime}{A}+2(P_r-P_\bot)\frac{R^\prime}{R}+\dot{q}
  \frac{B}{A}+2q\frac{B}{A}\left(\frac{\dot B}{B}+\frac{\dot R}{R}\right)=0.
\end{equation}
Using  (\ref{EE11}) and  (\ref{EE01})  we may write
\begin{equation}\label{EE11a}
  D_T U=\frac{m}{R^2}-4\pi R P_r+aE,
\end{equation}

\begin{equation}\label{21U}
  D_R\left(\frac{U}{R}\right)=\frac{4\pi q}{E}+\frac{\sigma}{R},
\end{equation}
which allows to rewrite (\ref{mupunto}) and (\ref{pprima}) as
\begin{eqnarray}\label{mupuntoU}
D_T \mu &+&\frac{1}{3}(3\mu +P_r+2 P_\bot)\Theta+\frac{2}{3}(P_r-P_\bot)\sigma+ED_R q\nonumber \\&+&2q\left(a+\frac{E}{R}\right)=0,
\end{eqnarray}
and

\begin{equation}\label{pprimaU}
 E D_R P_r+(\mu +P_r)a+2(P_r-P_\bot)\frac{E}{R}+D_T q+\frac{2}{3}q(2\Theta+\sigma)=0.
\end{equation}
Finally, combining  (\ref{EE11a}) with  (\ref{pprimaU}) we find
\begin{widetext}
\begin{equation}\label{pprimaU2}
  (\mu+P_r)D_T U=-(\mu+P_r)(4\pi P_r R^3-m)\frac{1}{R^2}-E^2 \left[D_R P_r+\frac{2}{R}(P_r-P_\bot)\right]
  -E\left[D_T q+\frac{2}{3}q(2\Theta+\sigma) \right].
\end{equation}
\end{widetext}
The above equation may be transformed further by replacing  (\ref{ecTra}) in  (\ref{pprimaU2}), and using  (\ref{EE11a})
\begin{widetext}
\begin{eqnarray}\label{pprimaU3}
 &&\left(\mu+P_r-\frac{\kappa T}{\tau}\right)D_T U = -\left(\mu+P_r-\frac{\kappa T}{\tau}\right)\left(4\pi R^3P_r-m\right) \frac{1}{R^2}
  - E^2\left [D_R P_r+\frac{2}{R}(P_r-P_\bot)-\frac{\kappa}{\tau}D_RT\right ]
 \nonumber \\ &+&Eq\left [\frac{1}{\tau}+\frac{1}{2}D_T\ln \left(\frac{\tau}{\kappa T^2}\right)-\frac{5}{6}\Theta-\frac{2}{3}\sigma\right ]
\end{eqnarray}
\end{widetext}

\end{document}